\documentclass[12pt]{article}
\usepackage{epsfig}
\usepackage{graphicx}
\usepackage{a4}
\begin{document}
\begin{titlepage}
\vspace{3 ex}
%
% ******************************* your title ************************
%
\begin{center}
{
\LARGE \bf \rule{0mm}{7mm}{\boldmath The CDFII Time-Of-Flight Detector and 
Impact on Beauty Flavor Tagging}\\
}
\end{center}

\vspace{4ex}
% Your name and Institution. Please do not include street addresses
% For example; University of California Los Angeles, California 90024, U.S.A.
% or Centre d'Etudes Nucleaires-Saclay, 91191 Gif-sur-Yvette, France.
% would be sufficient.

\begin{center}
C.Grozis$^{a}$, R.Kephart$^{a}$, R.Stanek$^{a}$, 
D.H.Kim$^{b}$, M.S.Kim$^{b}$, Y.Oh$^{b}$, 
Y.K.Kim$^{c}$, G.Vermendi$^{c}$, 
K.Anikeev$^{d}$, G.Bauer$^{d}$, 
I.K.Furic$^{d}$, A.Korn$^{d}$, I.Kravchenko$^{d}$, 
M.Mulhearn$^{d}$, Ch.Paus$^{d}$, S.Pavlon$^{d}$, K.Sumorok$^{d}$, 
C.Chen$^{e}$, M.Jones$^{e}$, 
W.Kononenko$^{e}$, J.Kroll$^{e}$, G.M.Mayers$^{e}$, M.Newcomer$^{e}$,
R.G.C.Oldeman$^{e}$, D.Usynin$^{e}$, R.Van Berg$^{e}$, 
G.Bellettini$^{f}$, C.Cerri$^{f}$, A.Menzione$^{f}$, E.Vataga$^{f}$, 
S.Dececco$^{g}$, D.Depedis$^{g}$,
C.Dionisi$^{g}$, 
\underline{S.Giagu$^{g,a}$,}
A.DiGirolamo$^{g}$, M.Rescigno$^{g}$, L.Zanello$^{g}$, 
S.Cabrera$^{h}$, J.Fernandez$^{h}$, G.Gomez$^{h}$, J.Piedra$^{h}$, T.Rodrigo$^{h}$,
A.Ruiz$^{h}$, I.Vila$^{h}$, R.Vilar$^{h}$, 
M.Ahn$^{i}$, B.J.Kim$^{i}$, S.B.Kim$^{i}$,
I.Cho$^{j}$, J.Lee$^{j}$, I.Yu$^{j}$, 
H.Kaneko$^{k}$, A.Kazama$^{k}$, 
S.Kim$^{k}$, K.Sato$^{k}$, K.Sato$^{k}$, F.Ukegawa$^{k}$

\vspace{1ex}
\small{\it{
$^{a}$Fermi National Accelerator Laboratory, USA \\
$^{b}$Kyungpook NAtional University, Korea \\
$^{c}$Lawrence Berkeley National Laboratory, USA \\
$^{d}$Massachusetts Institute of Technology, USA \\
$^{e}$University of Pennsylvania, USA \\
$^{f}$INFN, University of Pisa, Italy \\
$^{g}$INFN, University of Roma "La Sapienza", Italy \\
$^{h}$Instituto de F\'isica de Cantabria, Spain \\
$^{i}$Seoul National University, Korea \\
$^{j}$SungKyunKwan University, Korea \\
$^{k}$University of Tsukuba, Japan
}
}
\end{center}
\vspace{2 ex}
\vspace{2 ex}
%
% ************************* Your abstract ***************************
%
\begin{abstract}
The new CDFII detector incorporates a Time-of-Flight detector (TOF), 
employing plastic scintillator bars and fine-mesh photomultipliers.
Since August 2001 the TOF system has been fully instrumented and integrated 
into the CDFII data acquisition system.
With a design goal of $100~ps$ resolution the TOF system 
will provide at least two
standard deviations separation between $K^{\pm}$ and $\pi^{\pm}$ for
momenta $p<1.6~GeV/c$, complementing low momentum particle identification 
by means of the specific ionization energy loss measured in the 
drift chamber. 
We describe the design of the TOF detector and discuss 
the current status of its calibration and initial 
performances. Finally we review 
the expected impact of the TOF detector in the flavor tagging
of neutral $B_S$ meson.
\end{abstract}

\end{titlepage}

%*********************************************************
%
\clearpage

%\pagestyle{plain}
%\setcounter{page}{1}
%
% **************** Start of text **********************************
\section{Introduction}

Following the successful RunI from 1992 to 1996, the CDF detector has 
undergone a major upgrade~\cite{ref1} for the RunII which begun in March 2001.
The approval for the addition of a Time-of-Flight detector
was granted in January 1999. 
The installation of the TOF detector was completed in August 2001 
and its data has been included in the CDFII readout since then.

The primary physics motivation for TOF
is to complement and enhance the particle identification capability 
provided by the central drift chamber (COT) since it distinguishes 
$K^{\pm}$ and $\pi^{\pm}$ in the momentum region of their cross-over in 
$dE/dX$.
With an expected time-of-flight resolution of $100~ps$, the TOF system will 
be capable of identifying charged 
kaons from pions by their flight time difference with at least two standard 
deviation separation up to kaon momenta of $1.6~GeV/c$. 
Such an addition results in an enhancement of the $b$ flavor identification 
power, crucial to improve the statistical precision in CP violation 
and $B_s$ mixing measurements.

Particle identification with TOF is performed by measuring the
time of arrival of a particle
at the scintillator with respect to the collision time $t_0$. 
The particle mass $m$ can then be determined from the momentum $p$, 
the path-length $L$, measured by the tracking system, and the 
time-of-flight $t$ measured by the TOF:

\begin{equation}
   m = \frac{p}{c} \sqrt{\frac{c^2t^2}{L^2} - 1}
  \label{eq1}
\end{equation}

\begin{figure}[htb]
\begin{tabular}{cc}
\begin{minipage}[t]{0.5\linewidth}
\epsfig{file=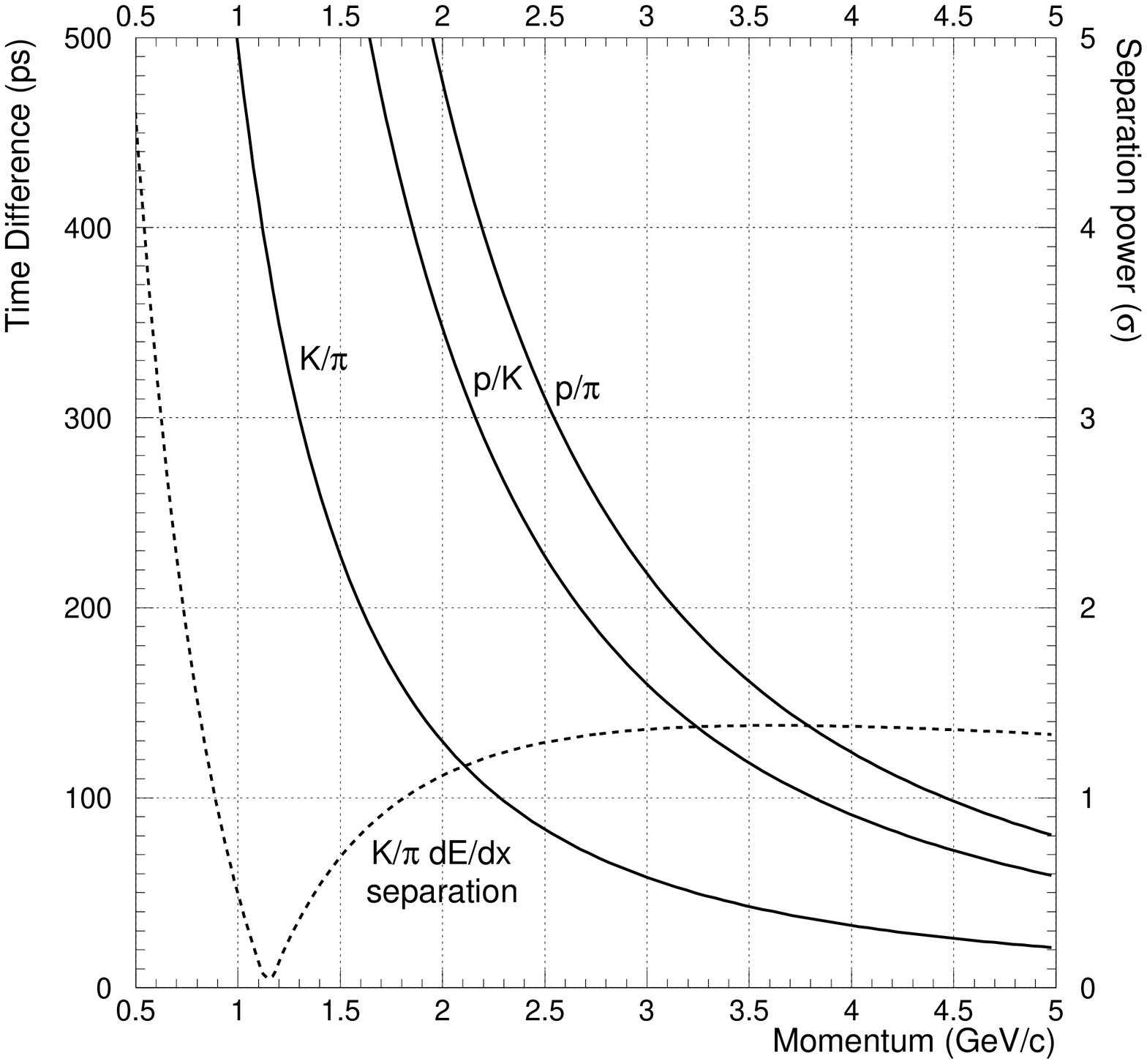, width=0.7\linewidth}
\caption{\small $K/\pi,p/\pi$ and $K/p$ time difference as a function of momentum 
over a path of $140~cm$, expressed in ps and separation power, assuming 
a resolution of 100 ps. The dashed line shows the $K/\pi$ separation power 
from $dE/dX$ measurement in the COT.}
\label{fig1}
\end{minipage}
&
\begin{minipage}[t]{0.5\linewidth}
\epsfig{file=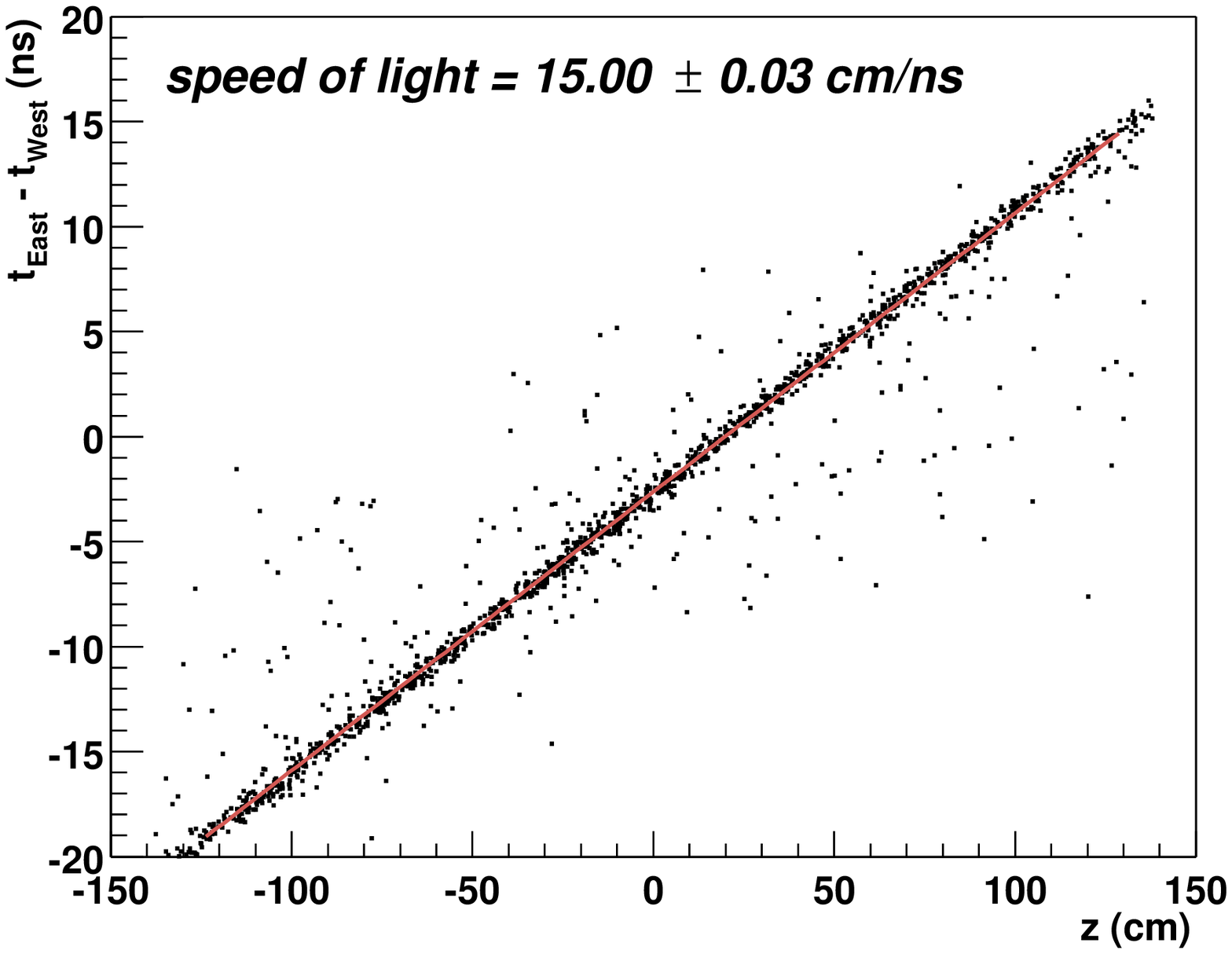,width=0.94\linewidth}
\caption{\small Distribution of the time difference, $t_{east} - t_{west}$, 
plotted versus the entrance point, $z$ of a track. The scattered points 
far away from the fitted line are due to multiple tracks hitting the bar.}
\label{fig2}
\end{minipage} \\
\end{tabular}
\end{figure}

Figure~\ref{fig1} shows the time-of-flight difference between $K/\pi$,
$p/K$ and $p/\pi$ and the separation power
assuming a resolution of $100~ps$. The TOF improves the $K/\pi$ separation 
in the momentum region of $p<1.6~Gev/c$ by $2\sigma$ or better.

\section{The CDFII TOF system}

The CDFII TOF detector~\cite{ref2},\cite{ref3},\cite{ref4}
 consists of 216 bars ($279~cm$ long and with a cross section 
of about $4x4~cm^2$) of Bicron BC-408 scintillator.
The bars are installed at a radius of $\sim 138~cm$ from the beam line
in the $4.7~cm$ of radial space available between the cylindrical 
outer wall of the central 
drift chamber (COT) and the inner wall of the cryostat of the 
super-conducting solenoid, 
covering a pseudo-rapidity region of roughly $|\eta|<1$.

A nineteen-stage fine mesh photomultiplier tube (PMT), 
Hamamatsu R7761, with 
a diameter of $1.5~$ inches is attached
to each bar end, for a total of $432$ PMTs in the system.
These operate in the $1.4~T$ solenoidal magnetic field 
with an average gain reduction of $500$ from the nominal gain of 
$10^6$.
The optical connection between the PMT and the scintillator is provided by 
a compound parabolic concentrator.  The PMT in turn is directly 
attached to a custom designed HV divider base which also 
connects to a preamplifier card for the readout.

Differential signal from the anode and the last dynode of 
the PMT is fed to the preamplifier, and drives, over $\sim12~m$ 
of shielded twisted pair cable, front-end electronics that reside in a VME 
crate mounted outside the detector. 
The front-end signal is split into two: one for timing measurement and the other for 
pulse height measurement.

The timing path enters a leading edge discriminator with an adjustable threshold, 
whose output serves as the start signal for the Time-to-Amplitude Conversion 
(TAC) circuit~\cite{ref5}. The TAC ramp is terminated by a common stop clock edge,
that is synchronized with the $p\bar{p}$ bunch crossing, and fanned 
out to all electronics channels with a design jitter less 
than $25~ps$.
The voltage output from the TAC is sampled by a 12-bit ADC, characterized 
by a least count of roughly $17~ps$ over a dynamic range of about $60~ns$.
TAC response has shown an excellent stability since the 
commissioning of the electronics, with a residual variation, after 
the calibrations, of less than $17~ps$.

The primary purpose for measuring the charge of PMT signals is to perform 
a correction for the dependence of the discriminator on signal amplitude.
The charge measurement is done by a charge sensitive ADC~\cite{ref5}.
The current driver is switched on by a 
gate of adjustable width, and is initiated by the discriminator output, so that 
only the charge due to the pulse that fired the discriminator is integrated.

\section{TOF Calibrations and Initial Performances}

The time at which a PMT pulse is registered, measured with respect 
to the nominal $p\bar{p}$ bunch crossing time, can be modeled by 
the following expression:

\begin{equation}
t = c + t_0 + tof + (L/2 \pm z)/s - S(Q)
\label{eq2}
\end{equation}
where $c$ is a constant offset describing the propagation 
delays in the cables, $t_0$ represents the time at which the 
$p\bar{p}$ interaction occurred, $tof$ is the time-of-flight of the 
particle impinging on the bar of scintillator of length $L$, $s$ 
is the effective speed of light in the scintillator,
and the last term describes the time-walk effect introduced by the
leading edge discriminator. The $z$ coordinate is measured 
along the length of the bar by the tracking chamber, the positive and negative 
sign of the term proportional to $z$ corresponds to a PMT in 
the west or east end of the detector, respectively.
The measurement of the parameters in eq.~\ref{eq2} for each channel 
is an essential step toward reaching the goal of $100~ps$ resolution, 
and it is performed regularly during periods of data taking.

When a single charged particle enters a bar, the time difference 
between signals reaching the east and west ends is essentially a linear function 
of the track entrance point along the bar, $z$. A typical distribution 
of the time difference plotted as a function of $z$ is shown in 
Figure~\ref{fig2}. The effective speed of light in the bar can be derived 
from the fitted slope. 
From the distribution of the residuals of this fit a first indication of
the timing resolution of each PMT, can be obtained. 
The resolution averaged over all bars is typically $250~ps$ or better. 
This is larger by approximately a factor two 
than the resolution of a mean time measurement, indicating that the TOF
resolution is not far from our design goals. 
It is worth notice here that as 
systematics effects can cancel in the time difference measurement,
this estimate represents only an indication of the capabilities of 
the detector.

Tracks which hit the scintillator at the same $z$ position 
are found to have an ADC response that is well 
described by a Landau distribution.
The width of the Landau peak divided by the peak position is typically 
$8-10\%$ for all channels. 
The dependence of the pulse height on the measured time and 
the parameterization of the time slewing effect
have been studied using a sample where each track passes through two 
adjacent bars of scintillator. Depending on the path length in each 
bar, a range of ADC responses is obtained on the two channels at the 
same end of the bar. At the same time because the track entrance 
point in each bar is similar, the time difference between the two channels 
depends mainly on time walk effects.
Due to finite attenuation length in the bars ($\sim 325~cm$) the 
pulse height depends on the entrance point of the track, $z$. For this 
reason the slewing correction introduces a linear dependence 
in the time difference $t_{east} - t_{west}$ versus $z$, resulting in a biased 
effective speed of light determination. 
After the time slewing correction the width of the speed of light 
distribution for all the PMTs is significantly reduced and observed low-side 
tail in the distribution disappeared.

\subsection{Preliminary Time Resolution Estimate and Particle ID Performances}

\begin{figure}[htb]
\begin{tabular}{cc}
\begin{minipage}[t]{0.5\linewidth}
\epsfig{file=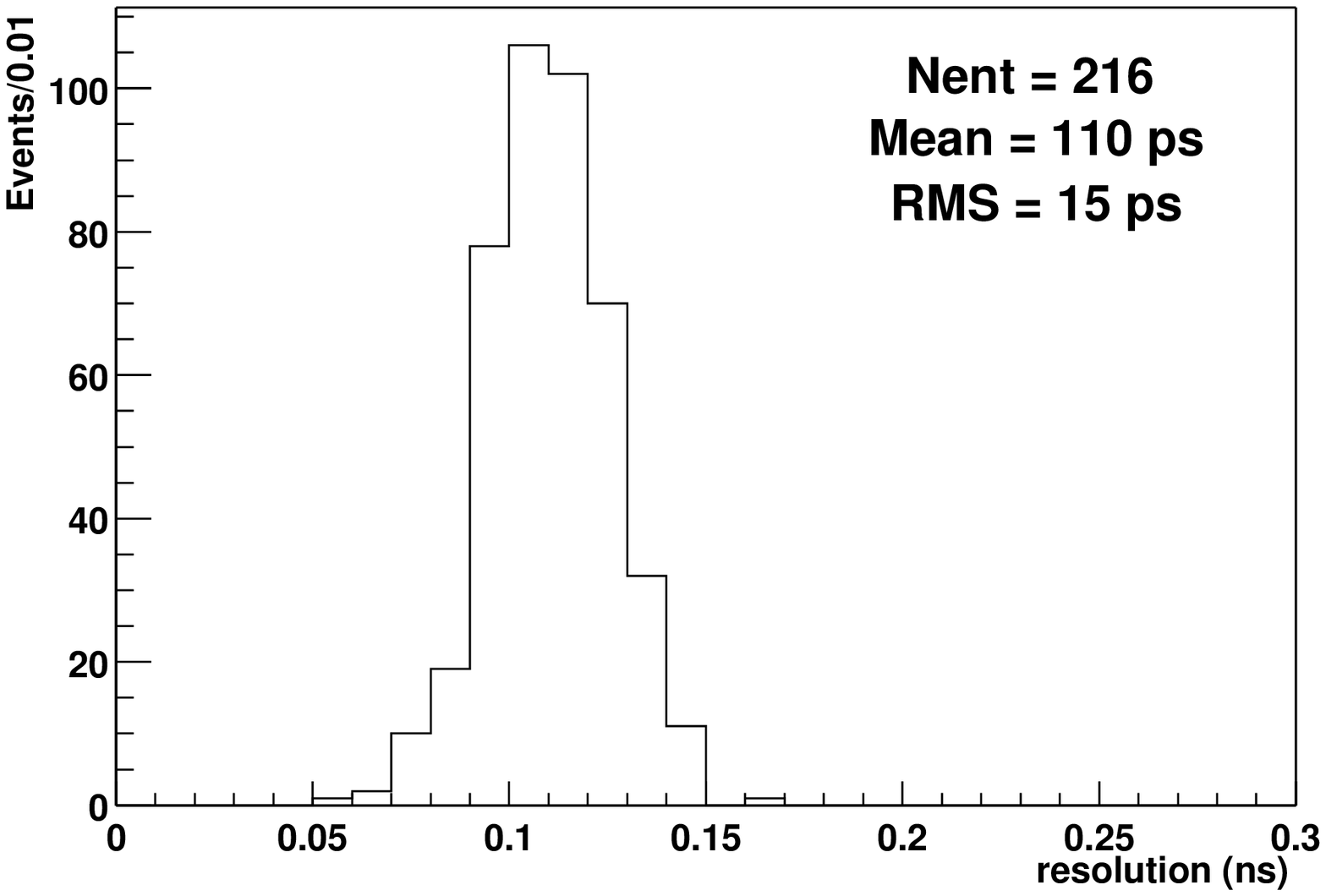,width=0.87\linewidth}
\caption{\small Time-of-flight resolution (at the PMT face) for all bars
of the CDFII TOF detector.}
\label{fig5}
\end{minipage}
&
\begin{minipage}[t]{0.5\linewidth}
\epsfig{file=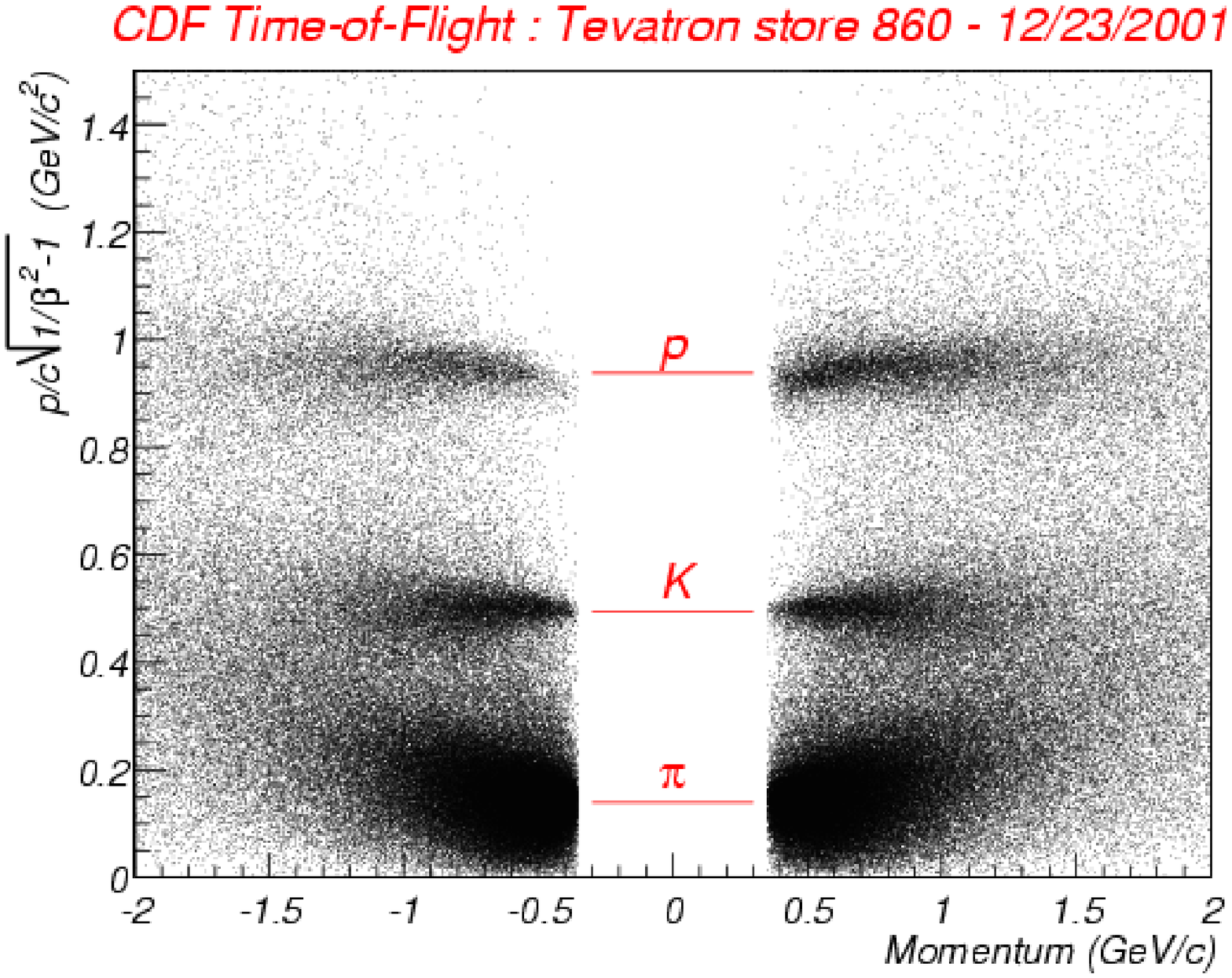,width=0.8\linewidth}
\caption{\small TOF reconstructed mass versus momentum for positive and 
negative tracks. The three horizontal lines correspond to nominal 
$p,K$ and $\pi$ masses.}
\label{fig6}
\end{minipage} \\
\end{tabular}
\end{figure}

After all calibrations and corrections were applied, the timing 
resolution of a given channel can be estimated by comparing 
the measured TOF for a track with the expected TOF under the pion mass 
hypothesis. 
Figure~\ref{fig5} shows the distribution of the resolution obtained 
with this procedure. It shows a mean value of 
$110~ps$ with an RMS of $15~ps$, which is close to the design goals.

In Figure~\ref{fig6} is shown the mass-momentum scatter plot computed using 
eq.~\ref{eq1} for low momentum tracks. Three clear clouds corresponding to 
pions, kaons and protons can be seen.

\begin{figure}[htb]
\center
\epsfig{file=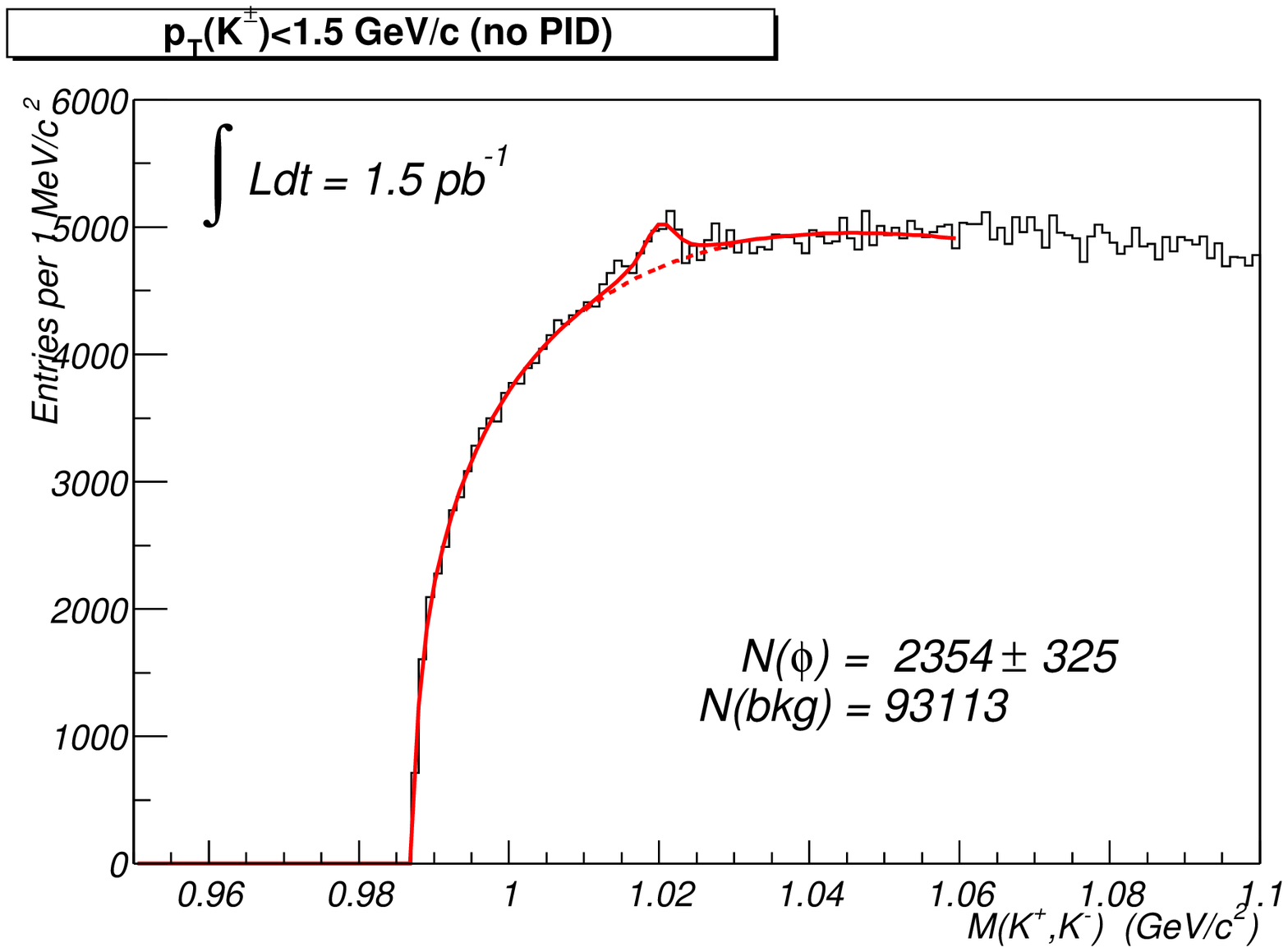,width=0.45\linewidth}\hbox{\hspace{1.0cm}} 
\epsfig{file=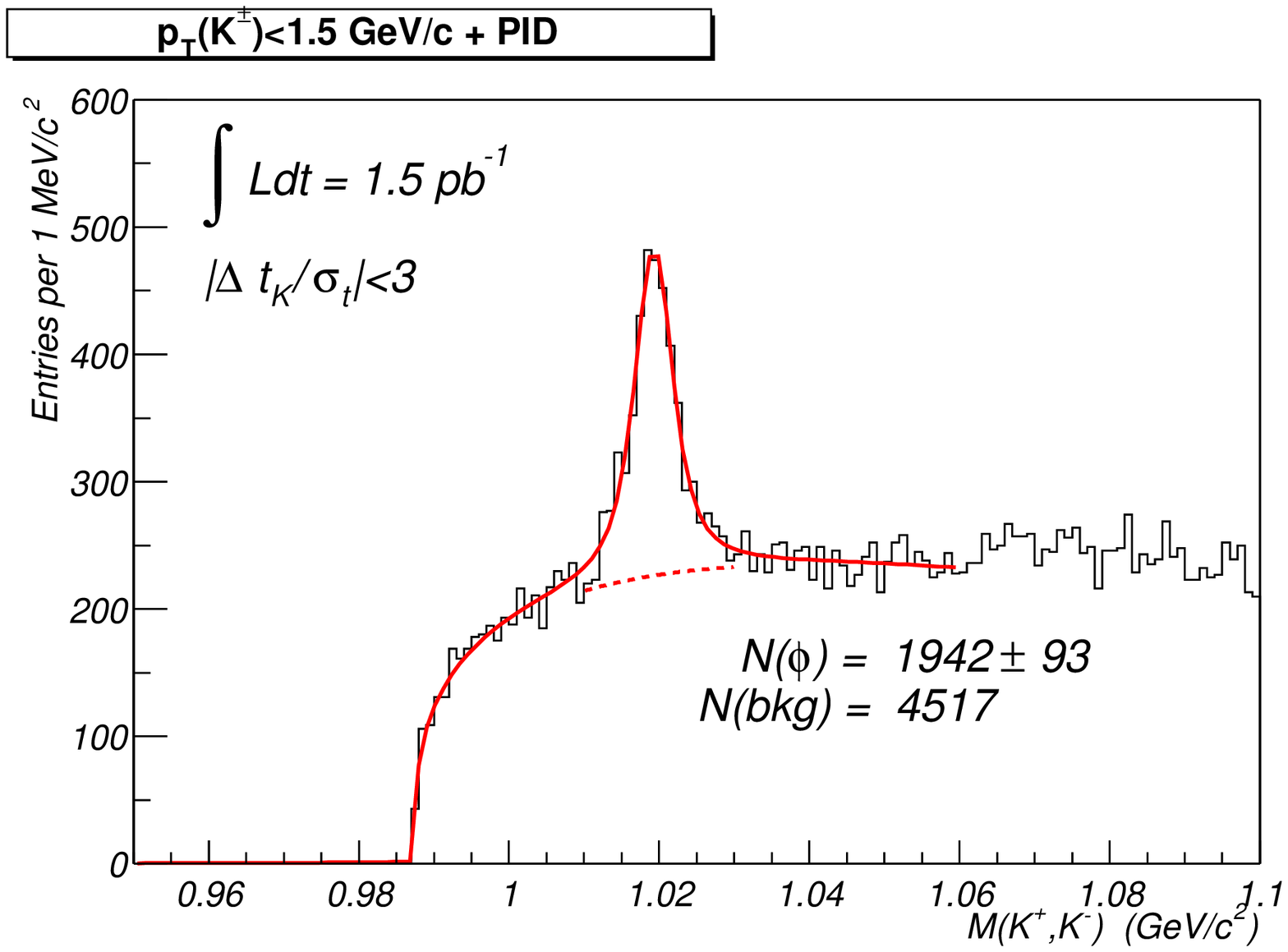,width=0.45\linewidth}
\caption{\small $K^+K^-$ invariant mass distribution for $p_T(K^\pm) < 1.5~GeV/c$, 
without (left) and with (right) TOF particle ID.}
\label{fig7}
\end{figure}

Actual realistic performance of the entire TOF system can be 
obtained by analyzing the $\phi \rightarrow K^+K^-$ signal in a sample of 
events recorded by CDFII. The $\phi$ candidates are selected 
from all pairs of oppositely charged tracks by assuming the kaon mass, 
with the only constraint that the two tracks originated from the same vertex. 
Both tracks are required to have a transverse momentum less than 
$1.5~GeV/c$ and valid reconstructed TOF informations. 
Figure~\ref{fig7} shows the invariant mass distributions obtained before 
and after applying TOF kaons particle identification. A clear 
signal peak in the distribution appears in the latter case, corresponding 
to a factor $20$ background reduction and $80\%$ signal efficiency.

\section{Flavor Tagging with TOF in CDFII}

The primary purpose of the TOF detector is to identify charged 
pions, kaons and protons. In the context of $B$ physics, particle 
identification is useful both for reconstructing $B$ decays and for 
$b$ flavor tagging.
In particular the CDF TOF detector was designed to substantially improve 
the statistical precision in the measurement of CP violation, as well 
as the sensitivity in measuring $B^0_S$ oscillations. \\
By the term "$b$ flavor tagging" we mean the ensemble of procedures 
and algorithms which act to determine the flavor of the $B$ meson when 
it was produced. 
The figure of merit for a $b$ flavor tagging method is its "total tagging 
effectiveness", $\epsilon D^2$. The efficiency $\epsilon$ accounts for 
the fact that we can apply 
a flavor tag only to a fraction $(0,1)$ of the $B$ candidates.
The dilution $D$ is related to the fact that sometime 
the obtained tag is wrong. If we call $P$ the probability that the tag is 
correct, the dilution is then defined as $D = 2P - 1$, so for a 
perfect tag we will have $D=1$, and for a random tag $D=0$.
The total tagging effectiveness $\epsilon D^2$ determines the effective statistical power
of the sample, 
so that, for example, the statistical error on the determination of a CP asymmetry 
$A$ with $N$ total candidates is proportional to
$\delta A \sim \sqrt{\frac{1}{\epsilon D^2 N}}$. An 
additional flavor tagging method that would increases $\epsilon D^2$ by a factor of two 
would lead to a reduced statistical error on $A$ by $\sqrt{2}$.

The methods of $b$ flavor tagging can be divided in two categories: 
"opposite-side"  tagging (OST)~\cite{ref7} and the "same-side" tagging (SST)~\cite{ref8}.
The former relays on the fact that the dominant production mechanism of $b$ 
quarks in hadron collisions generates $b\bar{b}$ pairs. To identify the flavor of 
a reconstructed $B$ meson of interest, we identify the flavor of the second $B$ hadron in the event 
and infer the flavor at production of the primary one. 
The same-side flavor tag exploits instead the correlation between $b$ 
flavor and the charge of the particles accompanying the reconstructed 
$B$. In particular, a $\pi^+$ is expected to be produced in association with
a $B^0$ meson, and a $\pi^-$ with a $\bar{B}^0$.

\begin{figure}[ht]
\begin{tabular}{cc}
\begin{minipage}[t]{0.5\linewidth}
\epsfig{file=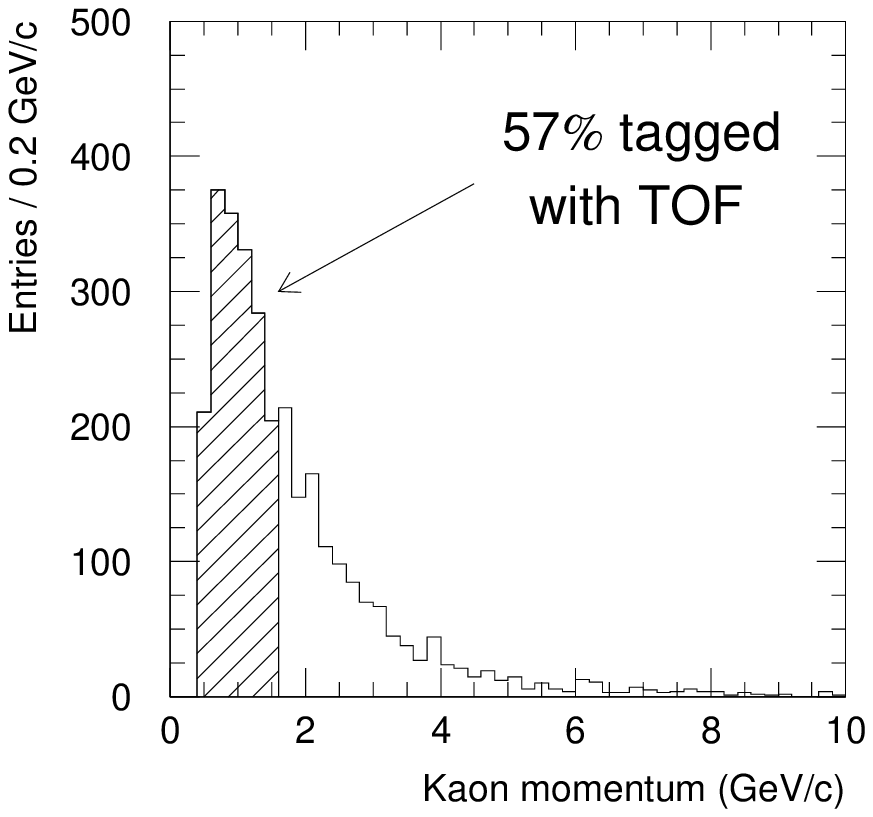,width=0.75\linewidth}
\caption{\small Momentum spectrum of kaons from $B$ decay opposite to
reconstructed $B^0\rightarrow J/\psi K^0_S$ decay obtained from
Montecarlo simulation. The region at $p < 1.6~GeV/c$, for 
which a $100~ps$ TOF resolution yields a better than 
$2\sigma$ $~K-\pi$ separation, is hatched.}
\label{fig8}
\end{minipage}
&
\begin{minipage}[t]{0.5\linewidth}
\epsfig{file=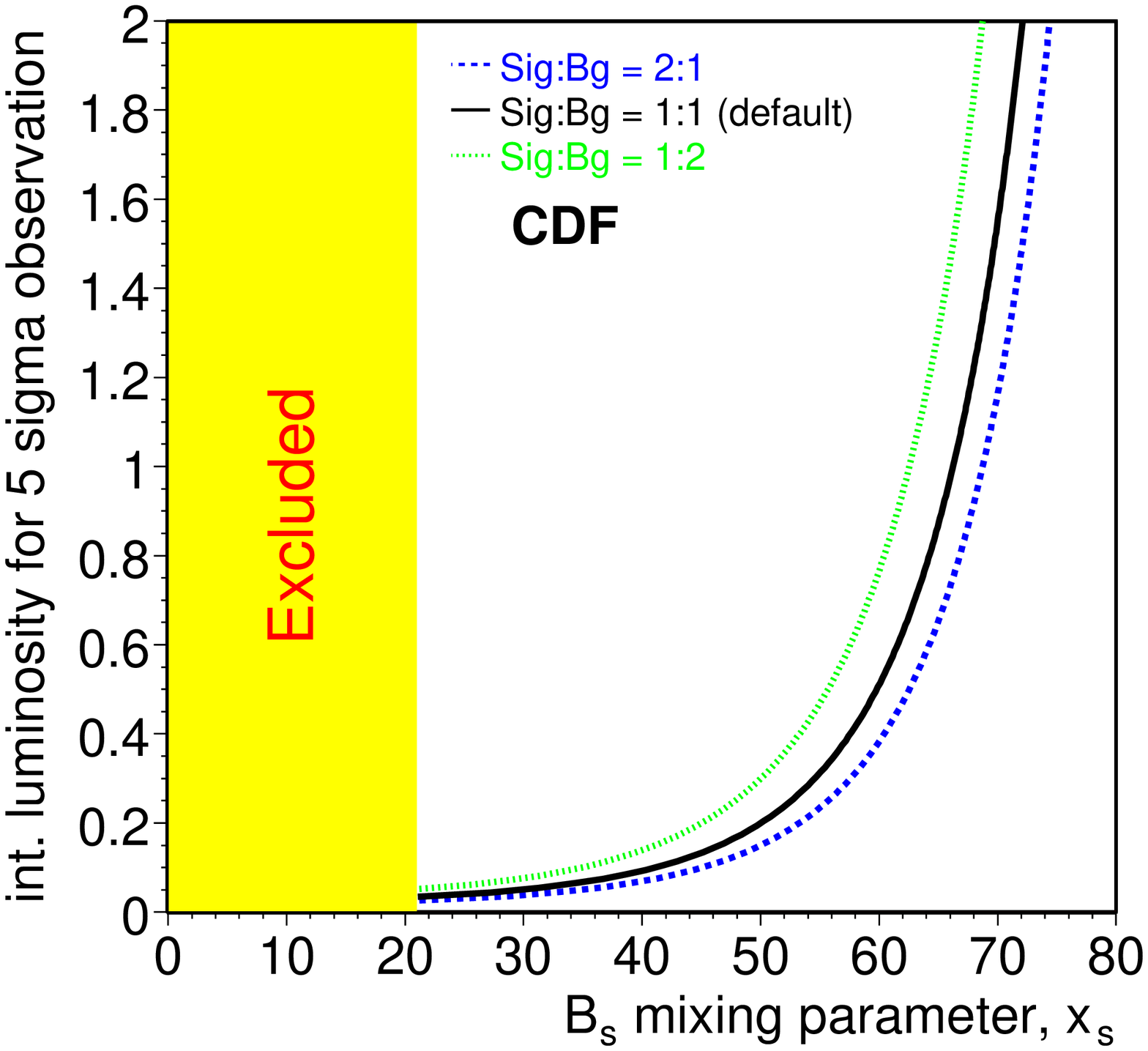,width=0.75\linewidth}
\caption{\small Luminosity required to achieve a 5 standard deviations observation 
of $B_S$ mixing as a function of the mixing parameter $x_S$, obtained using 
simulation and assuming a fully operational CDF detector.}
\label{fig9}
\end{minipage} \\
\end{tabular}
\end{figure}

The CDFII TOF can be used to enhance both opposite-side and same-side $b$ flavor 
tagging. The decays of $\bar{B}$ hadrons will contain more likely 
a $K^-$ than a $K^+$ in the final state due to $b \rightarrow c \rightarrow s$ week decays.
Therefore identifying a charged kaon from the $B$-decay can be used 
as an opposite-side $b$ flavor tag. 
In Figure~\ref{fig8} the Montecarlo momentum spectrum of kaons from 
the opposite-side $B$ decay to the $B^0/\bar{B}^0 \rightarrow J/\psi K^0_S$ signal 
is shown. More than $50\%$ of the available kaons are in a momentum range where 
the CDFII TOF can give a $2\sigma$ separation power from charged pions.
The CDFII potential for the opposite-side flavor kaon tag using the TOF has been 
estimated with a Montecarlo simulation, assuming the design resolution of 
$100~ps$. This gave $\epsilon D^2 = (2.4 \pm 0.2) \%$, which is 
more than a factor two higher than the typical values obtained 
in RunI~\cite{ref7},~\cite{ref8}.

The TOF can be used for same-side tagging as well.
In the hadronization process when a $B^0_S$ meson is produced an $s\bar{s}$ pair must 
be popped from the vacuum during fragmentation. The remaining $s$ or $\bar{s}$ 
quark can join a $\bar{u}$ or $u$ quark to form a charged kaon. 
The charge of the kaon thus depends on the flavor of the $B^0_S$ meson at production.
A factor four increase on $\epsilon D^2$ has been obtained, using Montecarlo simulations, 
for same-side tagging of $B^0_S$ using TOF informations (from $1.0\%$ to $4.2\%$).

Combining all the $b$ flavor tag methods available in RunII, CDFII has estimated 
a total $\epsilon D^2$ of $11.3\%$ for $B^0_S$ mesons using TOF for particle 
identification~\cite{ref9}. 
This large effectiveness will greatly increase the sensitivity for the measurement 
of the $B_S$ mixing parameter $x_S$, one of the unique measurement of 
Tevatron during the RunII. 
In Figure~\ref{fig9} the luminosity required to 
achieve a five standard deviation observation of mixing as a function of the mixing 
parameter $x_s$ is shown. 
The projection~\cite{ref9} is based on Montecarlo simulation of fully hadronic 
decays of the $B_S$ hadrons, and assumes a fully operational CDF detector and 
trigger system, work is in progress to determine the sensitivity using RunII data.
Within the Standard Model various experimental results indicate that 
$x_s \in (22.0, 30.8)$ at the $95\%$ confidence level~\cite{ref10}. 
Therefore, if mixing occurs as expected in the Standard Model, it should be 
observed even with a few hundreds of $pb^{-1}$ of total integrated luminosity.

\section{Conclusion}

The new CDFII Time-of-Flight detector become operational in
August 2001. Since then, it has working reliably without any significant 
problem. The first round of calibrations is available and preliminary 
studies show that the $100~ps$ time resolution goal seems reachable.
We expect substantial improvements using the TOF in the "B" physics 
program of CDF in coming years.

%***************** Bibliography ************************************
%\newpage

% **************** End of text ************************************

% **************** Your tables if any ************************************

%\newpage
%\clearpage

%\begin{table}
%%\centering
%\begin{tabular}{|r|c|c|}
%\hline 
%$\epsilon D^2$ & w/o TOF (\%) & with TOF (\%) \\\hline
%$B^0_S$        & $1.0 \pm 0.1$ & $4.2 \pm 0.3$ \\\hline
%$B^0$        & $1.8 \pm 0.1$ & $2.4 \pm 0.1$ \\\hline
%\end{tabular}
%\caption{\small Flavor tagging $\epsilon D^2$ values for same-side tagging of 
%$B^0_S$ and $B^0$ with and without TOF used for particle identification.}
%\label{tab1}
%\end{table}

%\begin{figure}[htb]
%\begin{tabular}{cc}
%\begin{minipage}[t]{0.45\linewidth}
%\epsfig{file=fig3.eps,width=0.97\linewidth}
%\caption{\small The ADC response, corrected for attenuation length effects, 
%fit to a Landau function.}
%\label{fig3}
%\end{minipage}
%&
%\begin{minipage}[t]{0.5\linewidth}
%\epsfig{file=fig4.eps,width=1.\linewidth}
%\caption{\small Effective speed of light distribution for all the bars before 
%(left) and after (right) the time slewing correction.}
%\label{fig4}
%\end{minipage} \\
%\end{tabular}
%\end{figure}

\end{document}